\documentclass[twoside,twocolumn,9pt]{article}
\usepackage{extsizes}
\usepackage[super,sort&compress,comma]{natbib} 
\usepackage[version=3]{mhchem}
\usepackage[left=1.5cm, right=1.5cm, top=1.785cm, bottom=2.0cm]{geometry}
\usepackage{balance}
\usepackage{times,mathptmx}
\usepackage{sectsty}
\usepackage{graphicx} 
\usepackage{lastpage}
\usepackage[format=plain,justification=justified,singlelinecheck=false,font={stretch=1.125,small,sf},labelfont=bf,labelsep=space]{caption}
\usepackage{float}
\usepackage{fancyhdr}
\usepackage{fnpos}
\usepackage[english]{babel}
\usepackage{array}
\usepackage{droidsans}
\usepackage{charter}
\usepackage[T1]{fontenc}
\usepackage[usenames,dvipsnames]{xcolor}

\newcommand{\davide}[1]{#1}

\newcommand{\dom}[1]{#1}
\usepackage{setspace}
\usepackage[compact]{titlesec}
\usepackage{amsmath}
\usepackage{amssymb}
\usepackage{amsfonts}
\usepackage{bm}

\usepackage{epstopdf}

\definecolor{cream}{RGB}{222,217,201}

\newcommand{\mean}[1]{\left\langle#1\right\rangle}
\newcommand{\minimizer}[1]{{#1}_\textnormal{min}}
\newcommand{\spinodal}[1]{{#1}_\textnormal{s}}
\newcommand{\critical}[1]{{#1}_\textnormal{c}}
\newcommand{\ddsp}[1]{\mathop{}\!\mathrm{d} #1 \mathop{}\!}

\begin{document}

\pagestyle{fancy}
\thispagestyle{plain}
\fancypagestyle{plain}{

\fancyhead[C]{\includegraphics[width=18.5cm]{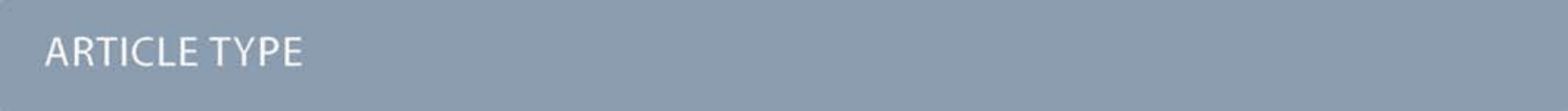}}
\fancyhead[L]{\hspace{0cm}\vspace{1.5cm}\includegraphics[height=30pt]{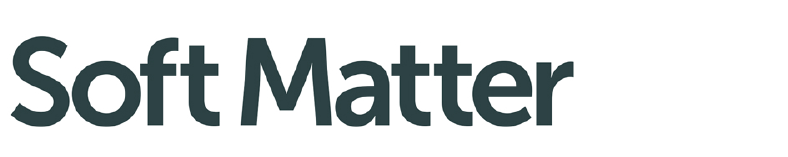}}
\fancyhead[R]{\hspace{0cm}\vspace{1.7cm}\includegraphics[height=55pt]{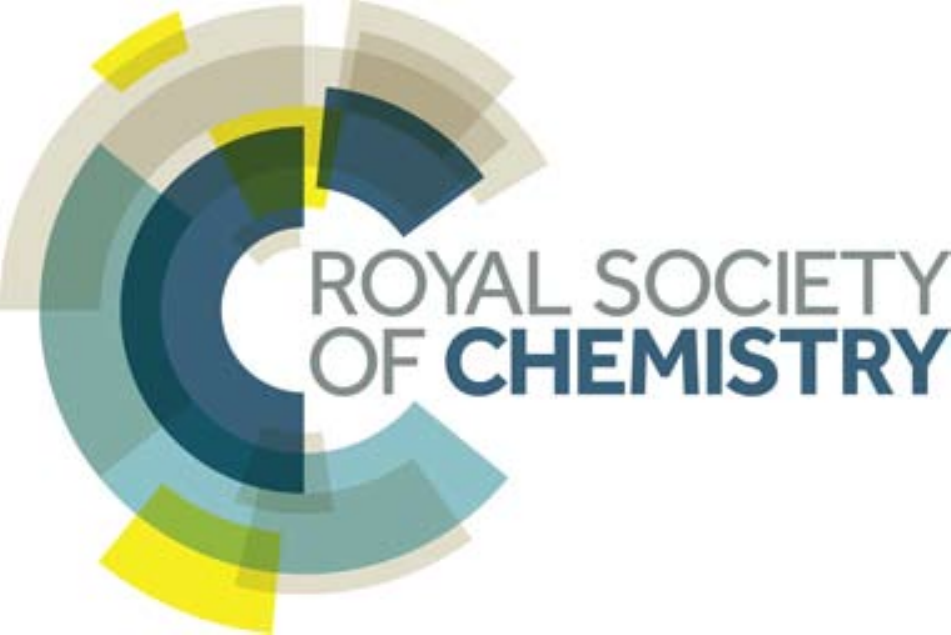}}
\renewcommand{\headrulewidth}{0pt}
}

\makeFNbottom
\makeatletter
\renewcommand\LARGE{\@setfontsize\LARGE{15pt}{17}}
\renewcommand\Large{\@setfontsize\Large{12pt}{14}}
\renewcommand\large{\@setfontsize\large{10pt}{12}}
\renewcommand\footnotesize{\@setfontsize\footnotesize{7pt}{10}}
\makeatother

\renewcommand{\thefootnote}{\fnsymbol{footnote}}
\renewcommand\footnoterule{\vspace*{1pt}%
\color{cream}\hrule width 3.5in height 0.4pt \color{black}\vspace*{5pt}} 
\setcounter{secnumdepth}{5}

\makeatletter 
\renewcommand\@biblabel[1]{#1}            
\renewcommand\@makefntext[1]%
{\noindent\makebox[0pt][r]{\@thefnmark\,}#1}
\makeatother 
\renewcommand{\figurename}{\small{Fig.}~}
\sectionfont{\sffamily\Large}
\subsectionfont{\normalsize}
\subsubsectionfont{\bf}
\setstretch{1.125} 
\setlength{\skip\footins}{0.8cm}
\setlength{\footnotesep}{0.25cm}
\setlength{\jot}{10pt}
\titlespacing*{\section}{0pt}{4pt}{4pt}
\titlespacing*{\subsection}{0pt}{15pt}{1pt}

\fancyfoot{}
\fancyfoot[LO,RE]{\vspace{-7.1pt}\includegraphics[height=9pt]{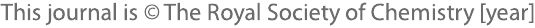}}
\fancyfoot[CO]{\vspace{-7.1pt}\hspace{13.2cm}\includegraphics{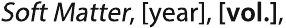}}
\fancyfoot[CE]{\vspace{-7.2pt}\hspace{-14.2cm}\includegraphics{RF}}
\fancyfoot[RO]{\footnotesize{\sffamily{1--\pageref{LastPage} ~\textbar  \hspace{2pt}\thepage}}}
\fancyfoot[LE]{\footnotesize{\sffamily{\thepage~\textbar\hspace{3.45cm} 1--\pageref{LastPage}}}}
\fancyhead{}
\renewcommand{\headrulewidth}{0pt} 
\renewcommand{\footrulewidth}{0pt}
\setlength{\arrayrulewidth}{1pt}
\setlength{\columnsep}{6.5mm}
\setlength\bibsep{1pt}

\makeatletter 
\newlength{\figrulesep} 
\setlength{\figrulesep}{0.5\textfloatsep} 

\newcommand{\topfigrule}{\vspace*{-1pt}%
\noindent{\color{cream}\rule[-\figrulesep]{\columnwidth}{1.5pt}} }

\newcommand{\botfigrule}{\vspace*{-2pt}%
\noindent{\color{cream}\rule[\figrulesep]{\columnwidth}{1.5pt}} }

\newcommand{\dblfigrule}{\vspace*{-1pt}%
\noindent{\color{cream}\rule[-\figrulesep]{\textwidth}{1.5pt}} }

\makeatother

\twocolumn[
  \begin{@twocolumnfalse}
\vspace{3cm}
\sffamily
\begin{tabular}{m{4.5cm} p{13.5cm} }

\includegraphics{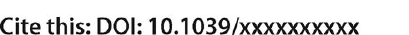} &



\noindent\LARGE{\textbf{Active turbulence and spontaneous phase separation in inhomogeneous extensile active gels}} \\

\vspace{0.3cm} & \vspace{0.3cm} \\

 & \noindent\large{Renato Assante\textit{$^{a}$}, Dom Corbett\textit{$^{a}$}, Davide Marenduzzo\textit{$^{a}$}, and  Alexander Morozov\textit{$^{a}$}} \\

\includegraphics{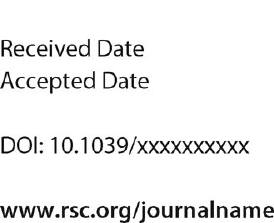} & \noindent\normalsize{
We report numerical results for the hydrodynamics of inhomogeneous lyotropic and extensile active nematic gels. By simulating the coupled Cahn--Hilliard, Navier--Stokes, and Beris--Edwards equation for the evolution of the \dom{composition}, flow and orientational order of an active nematic, we ask whether \dom{composition} variations are important to determine its emergent physics.
As in active gels \dom{of uniform composition}, we find that increasing either activity or nematic tendency (e.g., overall active matter \dom{concentration}) triggers a transition between an isotropic passive phase and an active nematic one. We show that \dom{composition} inhomogeneities are important in the latter phase, where we find three types of possible dynamical regimes. First, we observe regular patterns with defects and vortices: these exist close to the passive--active transition. Second, for larger activity, or deeper in the nematic phase, we find active turbulence, as in active gels \dom{of uniform composition}, but with exceedingly large \dom{composition} variation. In the third regime, which is \dom{uniquely associated with inhomogeneity} and occurs for large nematic tendency and \dom{low} activity, we observe spontaneous microphase separation into active and passive domains. The microphase separated regime is notable in view of the absence of an explicit demixing term in the underlying free energy which we use, and we provide a theoretical analysis based on the common tangent construction which explains its existence. 
We hope this regime can be probed experimentally in the future.}\\
\end{tabular}

 \end{@twocolumnfalse} \vspace{0.6cm}
]

\renewcommand*\rmdefault{bch}\normalfont\upshape
\rmfamily
\section*{}
\vspace{-1cm}

\footnotetext{\textit{$^{a}$~SUPA, School of Physics and Astronomy, The University of Edinburgh, James Clerk Maxwell Building, Peter Guthrie Tait Road, Edinburgh, EH9 3FD, United Kingdom; E-mail: dmarendu@ph.ed.ac.uk
}
}

\footnotetext{\dag~Electronic Supplementary Information (ESI) available: [details of any supplementary information available should be included here]. 
}



\section{Introduction}


Active gels~\cite{ramaswamy2010,prost2015,doostmohammadi2018} are fascinating examples of non-equilibrium soft matter. Some well-known realisations of these systems include solutions of cytoskeletal filaments, such as actin or microtubules, interacting with molecular motors, such as myosin or kinesin~\cite{mackintosh2010,sanchez2012}. Other instances come from living materials, and encompass microbial suspensions of algae or bacteria~\cite{hatwalne2004,wioland2013}. In an active gel, the constituent particles exert non-thermal forces on their environments. Such forces can be modelled, at the simplest level, as force dipoles, whose direction defines a nematic order parameter, which is a fundamental quantity to describe the emergent physics of these systems~\cite{marchetti2013}.

The activity arising from the distribution of force dipoles leads to a phenomenology which is strikingly different from that of passive colloidal particles or polymer suspensions. For instance, these materials harbour a ``spontaneous flow'' instability, which sets in for sufficiently strong activity, and comprises a non-equilibrium transition between a quiescent suspension and a state where activity fuels continuous motion~\cite{simha2002,voituriez2005,marenduzzo2007,marenduzzo2007b,cates2008}. For sufficiently large activity, the flow and orientation patterns of the spontaneously flowing state are seemingly chaotic, and the associated state is known as ``active turbulence''~\cite{giomi2015,doostmohammadi2017,stenhammar2017,bardfalvy2019,carenza2020,Kozhukhov2022}. In active turbulence, active gels self-organise into a random arrangement of vortices. Experiments and theories suggest that in the nematic phase these vortices have a typical length scale, arising from the competition between activity and elasticity~\cite{giomi2015}, while recent work points to important fundamental differences between active turbulence and its more widely studied passive counterpart~\cite{carenza2020,alert2020}. Active gels also possess strongly non-Newtonian rheological properties~\cite{saintillan2018}, such as marked activity-induced thinning or thickening~\cite{hatwalne2004,marenduzzo2007,cates2008,foffano2012b,martinez2020}, Darcy-like flow~\cite{mackay2020}, or negative drag in microrheology~\cite{foffano2012b,foffano2012}.

Existing theories and simulations of active gel hydrodynamics typically consider systems with constant \dom{composition}. In contrast, inspection of active turbulent patterns found with microtubule--kinesin mixtures in the presence of polyethylene-glycol (which causes adsorption to the oil--water interface~\cite{sanchez2012,martinez2019}) shows that the \dom{concentration} of active material is significantly inhomogeneous. While a linear stability analysis shows that in extensile gels, such as a microtubule--kinesin mixture, the onset of spontaneous flow depends on orientational bend fluctuations and \dom{compositional} fluctuations should be irrelevant~\cite{baskaran2009}, active turbulence is a highly non-linear phenomenon and the relevance of \dom{composition} inhomogeneities to its physics remains unclear. Additionally, passive colloidal particles aggregate in active nematics~\cite{foffano2019}, through a mechanism reminiscent of path coalescence~\cite{maritan1994,wilkinson2003} or fluctuation-dominated phase ordering~\cite{das2000}. This nonequilibrium aggregation shows that even a one-way coupling between \dom{composition} and spontaneous flow (as tracers are affected by the spontaneous flow but do not modify it) can in principle give rise to \dom{composition} inhomogeneities, and more in general to nontrivial physics outside the reach of a constant-\dom{composition} approximation.

To understand the role of \dom{compositional} inhomogeneities in the physics of spontaneous flow and active turbulence, here we study the hydrodynamic equations of motion of a mixture of an isotropic fluid and an active nematic gel by means of computer simulations. While the overall system is always incompressible, the active gel component can change \dom{its concentration}, as is realistic for the microtubule--kinesin mixtures considered in~\cite{martinez2019}. With respect to previous work on Cahn--Hilliard models coupled to active nematics focussing on the crossover between wet and dry systems driven by friction with the substrate~\cite{thampi2015,doostmohammadi2016,Thijssen2021}, here our focus is specifically on the qualitative role of \dom{compositional} inhomogeneities on the emerging physics and patterns. As in active gels \dom{of uniform composition}, we find that the system displays a transition between a passive isotropic phase and an active nematic phase. This transition can be triggered either by increasing activity or the nematic tendency of the system (more specifically the coupling between active matter \dom{concentration} and orientational order). In the active nematic phase, though, \dom{compositional} inhomogenities play a fundamental role and give rise to some unique dynamical behaviour. We find three dynamical regimes in this phase. Close to the transition boundary, our lyotropic system settles into regular flowing patterns, with approximately ordered spiral defect arrangements creating a rotational active flow consisting of long-lived and stable vortices. For larger activity, or deeper in the nematic phase, the flow becomes chaotic and we observe active turbulence. Both regular patterns and active turbulence are regimes which can be found in active gels \dom{of uniform composition}. There are differences though, as in our case the thermodynamic coupling between orientational and compositional order parameters favours a \dom{concentration} minimum, or relative void of active matter, at defect cores. Additionally, active turbulent patterns are characterised by a very broad distribution of local \dom{concentrations}. The last regime we observe is unique to inhomogeneous \dom{systems}, and is found even deeper in the nematic phase with respect to active turbulence, but for low activities. Here the active mixture spontaneously phase separates into low-\dom{concentration} disordered and high-\dom{concentration} nematic domains of irregular shape. We show by a semi-analytical theoretical analysis that this phase separation is due to the coupling between \dom{composition} and nematic ordering and hence is driven thermodynamically. The corresponding coarsening is arrested in our case by the spontaneous active flow, and the size of the steady state domains decreases with activity. 
We conclude by discussing ways in which our study can be taken forward, both theoretically and experimentally.

\section{Equations of motion}

To describe the equilibrium properties of an inhomogeneous active nematic system in the passive phase (i.e., when the activity parameter defined below is switched off), we employ a Landau--de Gennes free energy ${\cal F}$, whose density we call $f$. The latter consists of a contribution which characterises the orientational order of the active liquid crystal, measured by a nematic tensor $Q_{\alpha\beta}$, and of another contribution determining the physics of \dom{compositional} fluctuations, depending on the compositional order parameter $\phi$ (which measures the local \dom{concentration} of active material). The liquid crystalline free energy density we use is a standard one to describe passive nematic liquid crystals~\cite{degennes1993}, and is explicitly given by
\begin{align}
f_{\rm LC}=\frac{A_0}{2}\left(1 - \frac {\gamma(\phi)} {3}\right) Q_{\alpha \beta}^2 -
          \frac {A_0 \gamma(\phi)}{3} Q_{\alpha \beta}Q_{\beta
          \gamma}Q_{\gamma \alpha} \nonumber \\
+ \frac {A_0 \gamma(\phi)}{4} \left(Q_{\alpha \beta}^2\right)^2 + \frac{K}{2} \left(\partial_\gamma Q_{\alpha \beta}\right)^2,
\label{eqBulkFree}
\end{align}
where the first term is a bulk contribution, describing the isotropic--nematic transition, while the second term is an elastic distortion term. 
In the equation above, $A_0$ is a constant, $\gamma(\phi)$ controls the magnitude of order (so that it may be viewed as an effective temperature or concentration for thermotropic and lyotropic liquid crystals respectively), while $K$ is an elastic constant --- note we are using the (standard in this field) one-constant approximation~\cite{degennes1993}. Here and in what follows Greek indices denote cartesian components and summation over repeated indices is implied. The coupling between concentration and ordering arises through $\gamma(\phi)$, which equals
\begin{align}
\gamma(\phi) = \gamma_0 + \phi({\bf r},t) \Delta,
\label{eqCoupling}
\end{align}
where $\gamma_0$ and $\Delta$ are appropriate constants. In our simulations described below, we fix $\gamma_0$ and vary $\Delta$ (see Section 3 for full parameter list).

The free energy density used to describe \dom{compositional} fluctuations is instead given by a simple function, used to describe binary fluid in the mixed (non-phase-separating) regime, and its form is simply given by
\begin{align}\label{fphi}
f_{\phi}=\frac{a}{2}\phi^2, 
\end{align}
where $a$ is a constant related to the compressibility of the active gel component. Note we do not include a surface-tension-like square gradient term, $\frac{k_{\phi}}{2}\left(\partial_{\alpha}\phi\right)^2$, required for stabilisation in conventional binary mixture models, as \davide{density variations are already penalised thermodynamically by the term proportional to the elastic constant $K$ in Eq.~\ref{eqBulkFree}}. 
Note also that the total free energy density $f=f_{\rm LC}+f_{\phi}$ has two contributions which depend on $\phi$: besides the mixing free energy $f_{\phi}$, Eq.~\ref{fphi}, there is also a $\phi$ dependence in the liquid crystalline free energy $f_{\rm LC}$, Eq.~\ref{eqBulkFree}, through the $\gamma(\phi)$ term. 

The fluid velocity, ${\bf u}$, obeys the continuity equation and the Navier--Stokes equation,
\begin{align}
\label{incompressibility}
\partial_\alpha u_\alpha &= 0, \\
\label{Stokes}
\rho\left(\partial_t+ u_\beta \partial_\beta\right)
u_\alpha &= -\partial_\alpha p_0 + \eta \partial_\beta^2 u_\alpha +  \partial_\beta \Pi_{\alpha\beta} -\zeta \partial_\beta \left(\phi Q_{\alpha\beta}\right),
\end{align}
where $\rho$ is the fluid density, $\eta$ is an isotropic viscosity, and 
\begin{align}
&\Pi_{\alpha\beta}= 2\xi
(Q_{\alpha\beta}+{1\over 3}\delta_{\alpha\beta})Q_{\gamma\epsilon}
H_{\gamma\epsilon} \\ \nonumber
&-\xi H_{\alpha\gamma}(Q_{\gamma\beta}+{1\over
  3}\delta_{\gamma\beta})-\xi (Q_{\alpha\gamma}+{1\over
  3}\delta_{\alpha\gamma})H_{\gamma\beta} \nonumber \\
&-\partial_\alpha Q_{\gamma\nu} {\partial
f\over \partial\partial_\beta Q_{\gamma\nu}}
+Q_{\alpha \gamma} H_{\gamma \beta} -H_{\alpha
 \gamma}Q_{\gamma \beta}.
\label{BEstress}
\end{align}
is the stress tensor, where $\zeta$ is the activity~\cite{marchetti2013}, and measures the strength of active force dipoles. With the sign convention chosen here $\zeta>0$ means extensile rods and $\zeta<0$ means contractile ones~\cite{marchetti2013}. The molecular field ${\bf H}$ which provides the driving motion is given by
\begin{equation}
{\bf H}= -{\delta {\cal F} \over \delta {\bf Q}}+({\bf
    I}/3) {\rm Tr} {\delta {\cal F} \over \delta {\bf Q}},
\label{molecularfield}
\end{equation}
where {\rm Tr} denotes the tensorial trace. 

The equation of motion for {\bf Q} is taken to be \cite{beris1994}
\begin{align}
\partial_t {\bf Q} + {\bf u}\cdot {\nabla \bf Q} = \Gamma {\bf H} + {\bf S} 
\label{Qevolution},
\end{align}
where $\Gamma$ is a collective rotational diffusion constant.
The first term on the left-hand side of Eq.~\ref{Qevolution}
is the material derivative describing the usual time dependence of a
quantity advected by a fluid with velocity ${\bf u}$. This is
generalized for rod-like molecules by a second term
\begin{align}
{\bf S} = \left(\xi  {\bf D} + {\bf \omega}\right)\cdot\left({\bf Q} + \frac{1}{3}{\bf I}\right) + \left({\bf Q} + \frac{1}{3}{\bf I}\right)\cdot \left(\xi  {\bf D} - {\bf \omega}\right) \nonumber \\
- 2\xi \left({\bf Q} + \frac{1}{3}{\bf I}\right)\,{\rm Tr}\left( {\bf Q}\cdot\nabla{\bf u}\right),
\label{S_definition}
\end{align}
where
\begin{align}
{\bf D} = \frac{\nabla {\bf u} + \nabla {\bf u}^\dagger}{2}, \\ \nonumber
{\bf \omega} = \frac{\nabla {\bf u} - \nabla {\bf u}^\dagger}{2},
\end{align}
are the symmetric part and the anti-symmetric part respectively of the velocity gradient tensor $\partial_\beta u_\alpha$.
The constant $\xi$ depends on the molecular details of a given liquid crystal. The first term on the right-hand side of Eq.~\ref{Qevolution} describes the relaxation of the order parameter towards the minimum of the free energy. 

Finally, the active material \dom{concentration}, $\phi$, obeys a Cahn--Hilliard-like equation,
\begin{align}
\partial_t {\bf \phi} + {\bf u}\cdot {\nabla \phi} = M \nabla^2 \mu,
\label{phievolution}
\end{align}
where $\mu=\frac{\delta{\cal F}}{\delta \phi}$ is the chemical potential of the active mixture, and $M$ is a mobility, which for simplicity we consider to be constant. 

\if{To characterise the local nature of the flow --- i.e., to determine to what extent it is locally extensional, shear or rotational --- we start from the tensors ${\bf D}$ and ${\bf \omega}$ which we have just defined, and calculate the following invariants,
\begin{align}
|{\bf D}| = \sqrt{\frac{1}{2}\left({\bf D}:{\bf D}\right)}=\sqrt{\frac{1}{2}\sum_{i,j}D_{ij}^2}
\end{align}
and
\begin{align}
|{\bf \omega}| = \sqrt{\frac{1}{2}\left({\bf \omega}:{\bf \omega}\right)}=\sqrt{\frac{1}{2}\sum_{i,j}\omega_{ij}^2}.
\end{align}
We then define a flow parameter $\chi$ as 
\begin{align}
\chi=\frac{|{\bf D}|-|{\bf \omega}|}{|{\bf D}|+|{\bf \omega}|}.
\end{align}
For purely rotational, shear and extensional flow the flow parameter attains the values $\chi=-1$, $0$ and $1$ respectively, independent of the coordinate system chosen to describe the system. As $\chi$ can be defined locally, we use it below to characterise the nature of the local spontaneous flow in active nematics.}\fi

\section{Numerical method}

To study the dynamics of Eqs.(\ref{incompressibility})-(\ref{phievolution}), here we perform direct numerical simulations. Note that in our simulations we assume that the fields are two-dimensional and that the ${\bf Q}$ tensor describes nematic order in a 2D plane (i.e., we assume that there is no out-of-plane nematic order). We employ an in-house MPI-parallel code developed within the Dedalus spectral framework \cite{burns2020}. Simulations are performed on a periodic rectangular domain $[0,H]\times[0,H]$ with $H=200$ (here and below, all quantities are given in simulation units). All fields are represented by de-aliased, double periodic Fourier series with $512 \times 512$ Fourier modes. Our time-iteration scheme employs a 4th-order semi-implicit backward differentiation formula (BDF) scheme \cite{Wang2008} with the timestep $dt=0.1$. We have confirmed that our spatial and temporal accuracy is sufficient to obtain numerically converged results. To obtain statistically converged averages, simulations are performed for $30000$ time units.

In what follows, we fix $\rho=2$, $\eta=5/3$, $\xi=0.7$, $A_0=0.1$, $K=0.01$, $\gamma_0=2$, $\Gamma=1$, and $a=0.003$, and $M=4$. These parameters are chosen as they are in line with those used in previous hybrid lattice Boltzmann simulations of both constant-\dom{composition} active nematics~\cite{marenduzzo2007,marenduzzo2007b} and self-motile active gel droplets~\cite{tiribocchi2015}.

\section{Results}

To discuss our results, we first present numerical simulations, and then a semi-analytical discussion of phase separation in the passive limit which sheds light on the phase diagram which we find. 

\subsection{Simulation results}

To address the role of \dom{compositional} inhomogeneities in the hydrodynamics of extensile active gels, we first report numerical simulations (for methodology details, see Section 3). 
We characterise the dynamical behaviour of the system as a function of two key parameters: activity, quantified by $\zeta$, and tendency to acquire nematic order, quantified by $\gamma(\phi_0)=\gamma_0+\phi_0\Delta$. In practice, we change $\zeta$ and $\Delta$ in our simulations, keeping other parameters fixed (see Section 3 for a full list). 

For each set of parameter values, we compute: (i) the average largest eigenvalue of ${\bf Q}$, denoted by $\langle q \rangle$, which we use to quantify the global magnitude of order; (ii) the average fluid velocity, $\langle |{\bf u}| \rangle$; and (iii) the average variance of the compositional order parameter. In each case, these averages are computed first spatially, over a configuration, and then over time. We acquire data when the system is in a statistical steady state, removing any initial transient. These quantities allow us to build a phase diagram.

\begin{figure}[!h]
  \includegraphics{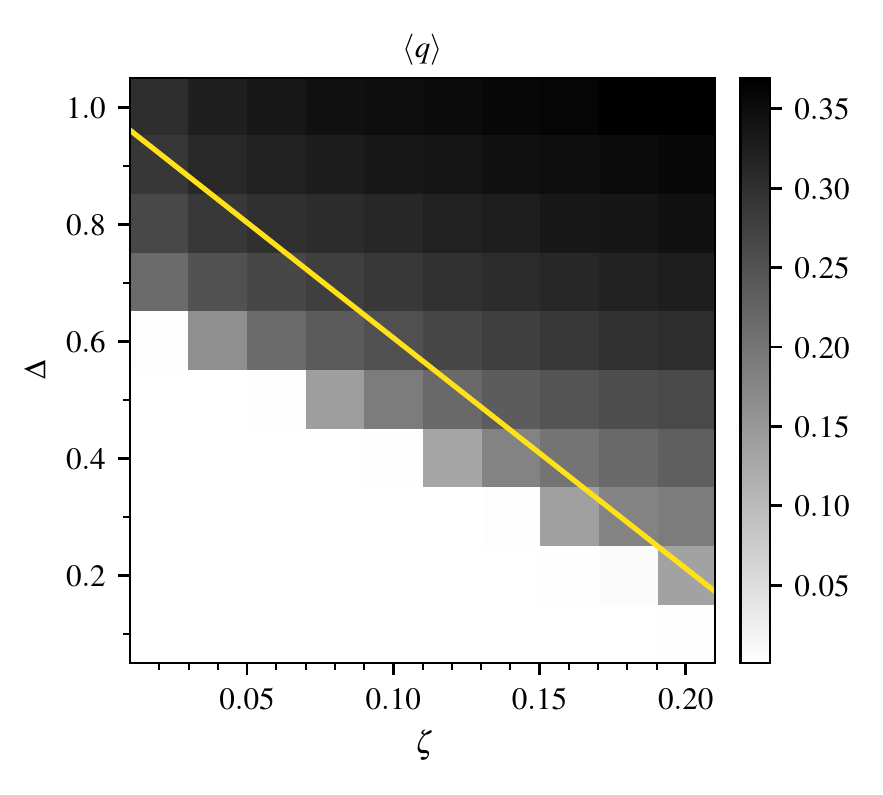}
  \caption{Heatmap of the magnitude of nematic order as a function of $\zeta$ and $\Delta$. The plot can be used to define regions in parameter space where the system is in the isotropic passive (bottom left) or active nematic (top right) phase. The prediction for the onset of spontaneous flow (passive--active transition) from linear stability analysis (see text) is shown as a yellow line. Note that the isotropic--nematic transition in the passive limit ($\zeta=0$) occurs at $\gamma(\phi_0)=2.7$ (corresponding to $\Delta=0.7$; cf.\ section~\ref{subsec:passive-demixing}).}
\label{fig1}
\end{figure}

The magnitude of order, $\langle q \rangle$, is shown in Fig.~\ref{fig1} and allows us to map the boundary between the isotropic and nematic phase. The nematic phase can be reached by increasing either $\Delta$ (thereby $\gamma(\phi_0)$) or $\zeta$. The former is a thermodynamic route, the latter is a non-equilibrium one. The non-equilibrium transition between isotropic and nematic phases arises because activity-induced flows create shear, which in turn generates nematic order. Analogous transitions in systems with \emph{constant} \dom{composition} or polar order have been studied in~\cite{thampi2015,santhosh2020,giordano2021}. A linear stability calculation analogous to that of~\cite{santhosh2020} shows that an isotropic state of constant $\phi$ is unstable, for sufficiently large $\zeta$, to an active nematic state with non-zero flow and order. The critical threshold in the $\left(\Delta,\zeta\right)$ plane is described by $\zeta_\textnormal{c}(\Delta)$, where
\begin{equation}\label{linearstabilityanalysis}
    \zeta_\textnormal{c}(\Delta)=\frac{2}{3}\xi A_0 \left(1-\frac{\gamma_0+\phi_0\Delta}{3}\right)\left(1+\frac{9\Gamma\eta}{2\xi^2}\right),
\end{equation}
corresponding to a straight line on the $(\Delta,\zeta)$ plane.
This predicted boundary is shown as a yellow line in Fig.~\ref{fig1}. It agrees well with our numerics for small $\Delta$, sufficiently far from the passive isotropic--nematic transition (at $\zeta=0$ and $\Delta=0.7$). As the latter is a first-order discontinuous transition,~\davide{which requires the inclusion of non-linear terms in the equation of motion to be accurately described,} it is unsurprising that there is a quantitative discrepancy with our linearised calculation close to this point.

\begin{figure*}[!h]
  \includegraphics{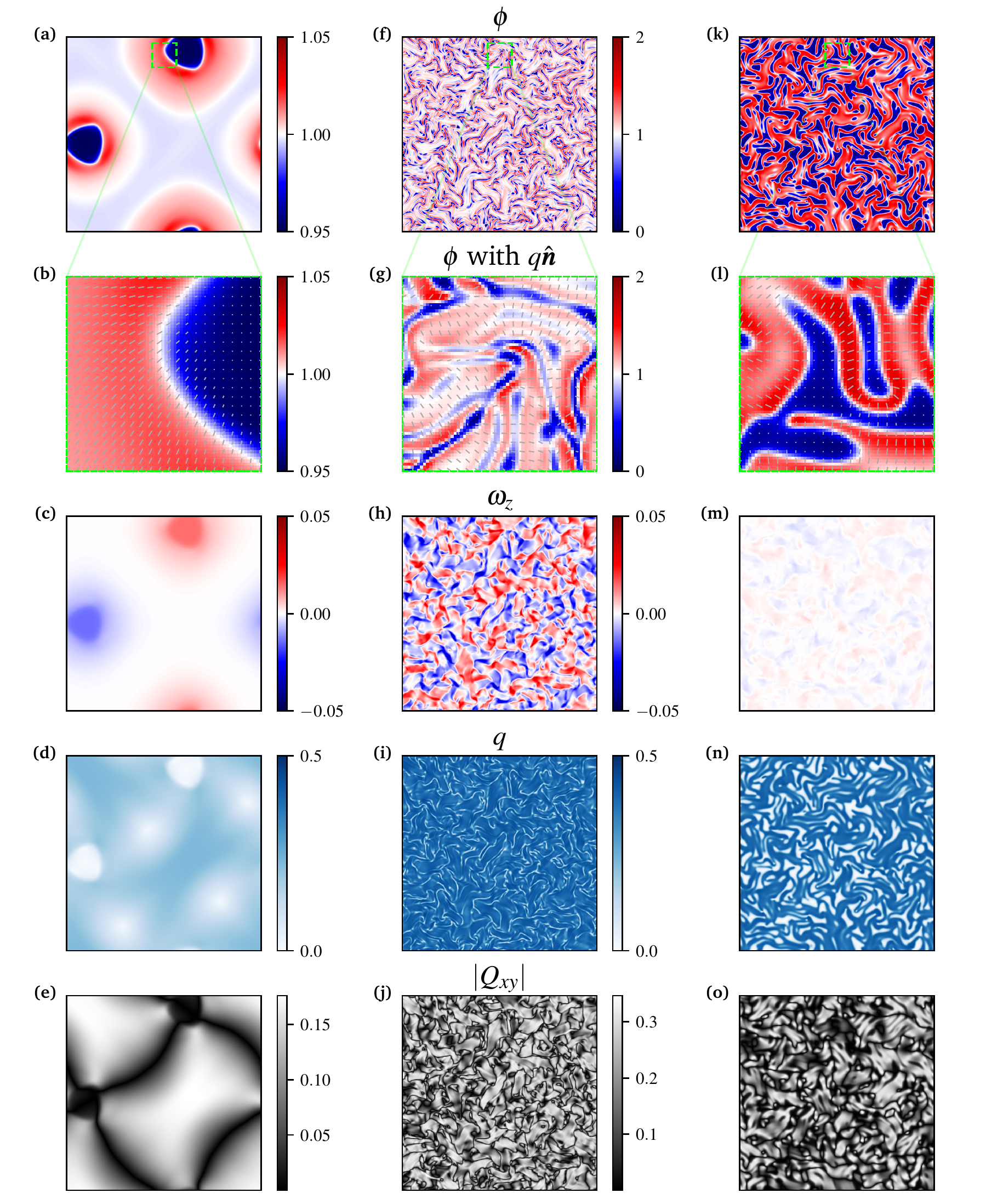}
  \caption{From top to bottom rows represent: snapshots of \dom{concentration} field, $\left\{\phi(x, y) \mid x, y \in [0,200]\right\}$; director field, $\bm{\hat{n}}$ (scaled by local magnitude of order, $q$\dom{, and coarse grained for visibility}), overlaid on \dom{concentration} (for $x \in [87, 113]$, $y \in [168, 194]$); vorticity ($\omega_z$); largest eigenvalue of $\bm{Q}$ tensor ($q$); $|Q_{xy}|$, qualitatively corresponding to a Schlieren pattern as could be obtained experimentally with crossed polarisers. 
  Parameters are as in Section except: (a-e) $\zeta=0.16$, $\Delta=0.3$; (f-j) $\zeta=0.2$, $\Delta=1$; (k-o) $\zeta=0.02$, $\Delta=1$. Colour scales are shared between (f)-(j) and (k)-(o).}
\label{fig2}
\end{figure*}

Throughout the nematic phase, except at $\zeta=0$, we find a non-zero flow in steady state. In other words, the emerging nematic structures are always active in the parameter range which we have explored. To characterise their behaviour, we show in Fig.~\ref{fig2} the dynamical regimes we observe, before discussing the phase diagram quantitatively. 

Fig.~\ref{fig2}(a--e) shows an example of regular active patterns. These are found close to the isotropic--nematic transition for moderate values of $\Delta$. The example shown features regular \dom{compositional} modulations in steady state (Fig.~\ref{fig2}a), where local minima are \dom{collocated with} spiral defects \dom{in the director field} of topological charge $1$. The latter are most easily visible through the 4-brush pattern in the Schlieren-like plot of $|Q_{xy}|$ in Fig.~\ref{fig2}e~\cite{degennes1993}, and also correspond to deep minima in the local nematic order parameter, as shown in Fig.~\ref{fig2}d. There are also steady vortices associated with the pattern, because spirals continuously rotate as in previous models of defects in constant-\dom{composition} active nematics~\cite{kruse2004}. Such vortices can be identified as maxima and minima in the vorticity plot in Fig.~\ref{fig2}c. \dom{This collocation with vortices provides support for the interpretation that these thermodynamically unstable structures are stabilised in steady state by the active flow.}
It is the thermodynamic coupling (proportional to $\Delta$) between order parameter and \dom{concentration} in our free energy that drives the local \dom{concentration} depletion at the centres of spirals, giving rise to a correlation between \dom{concentration} and order, where larger \dom{concentration} is associated with larger order, and vice versa. \dom{Thus the elastic energy cost associated with defect formation is decreased through proximal depletion, which explains the collocation of defects and voids.}
Our simulations show that the regular defect patterns we find close to the transition can either be stationary or self-motile (see Suppl.\ Movie~\dom{2} for the dynamics correspondent to the snapshot in Fig.~\ref{fig2}a--e). 

Fig.~\ref{fig2}(f--j) shows an example of a different dynamical regime, obtained deeper in the nematic phase, for sufficiently large $\zeta$ and $\Delta$ (see also Suppl.\ Movie~\dom{3}). Here, the patterns are never stable but display a chaotic dynamics and diffuse around in the system. In line with constant-\dom{composition} active nematic literature~\cite{doostmohammadi2017,carenza2020,alert2020}, we refer to these spatiotemporally varying patterns as active turbulence. Our simulations show that active turbulent patterns are accompanied by the appearance of defects --- mostly of half-integer topological charge, corresponding to two-brush patterns in $|Q_{xy}|$ (Fig.~\ref{fig2}j) --- as in active gels \dom{of uniform composition}. We also find this regime is characterised by large \dom{concentration} variations (Fig.~\ref{fig2}f).
As in the case of regular patterns, here too the \dom{concentration} field tends to decrease close to defect cores (Fig.~\ref{fig2}f,g). As the active chaotic spontaneous flow moves defects around, streaks of \dom{concentration} voids form (Fig.~\ref{fig2}f,g) which follow defect trajectories.

While the regimes in Figs.~\ref{fig2}(a--e) and \ref{fig2}(f--j) are qualitatively reminiscent of those found in the literature for constant-\dom{composition} active extensile gels~\cite{marenduzzo2007,thampi2015,santhosh2020}, we also find a third dynamical regime which is unique to inhomogeneous \dom{systems}. This regime is found for low activity and sufficiently large $\Delta$, and consists in spontaneous phase separation into active and passive domains (Fig.~\ref{fig2}(k--o), see also Suppl.\ Movie~\dom{1}). Active domains are nematically ordered. 
Importantly, this is an example of microphase separation, or arrested phase separation, as coarsening does not proceed indefinitely and there are multiple domains in steady state \davide{(Figs.~\ref{fig2}k,l, and S1). In other words, the late-time domain size --- computed, for instance, via the inverse first moment of the structure factor ---  does not scale with system size and decreases with increasing $\zeta$ (see Supplemental Material and Figs.~S1, S2).} 
The presence of this spontaneous microphase separation regime is at first sight surprising, as the $\phi$-dependent part of the free energy does not promote demixing by itself. As we show in section~\ref{subsec:passive-demixing}, however, phase separation is driven by the coupling between nematic order and local \dom{concentration} of active matter. 

\begin{figure}[!h]
  \includegraphics{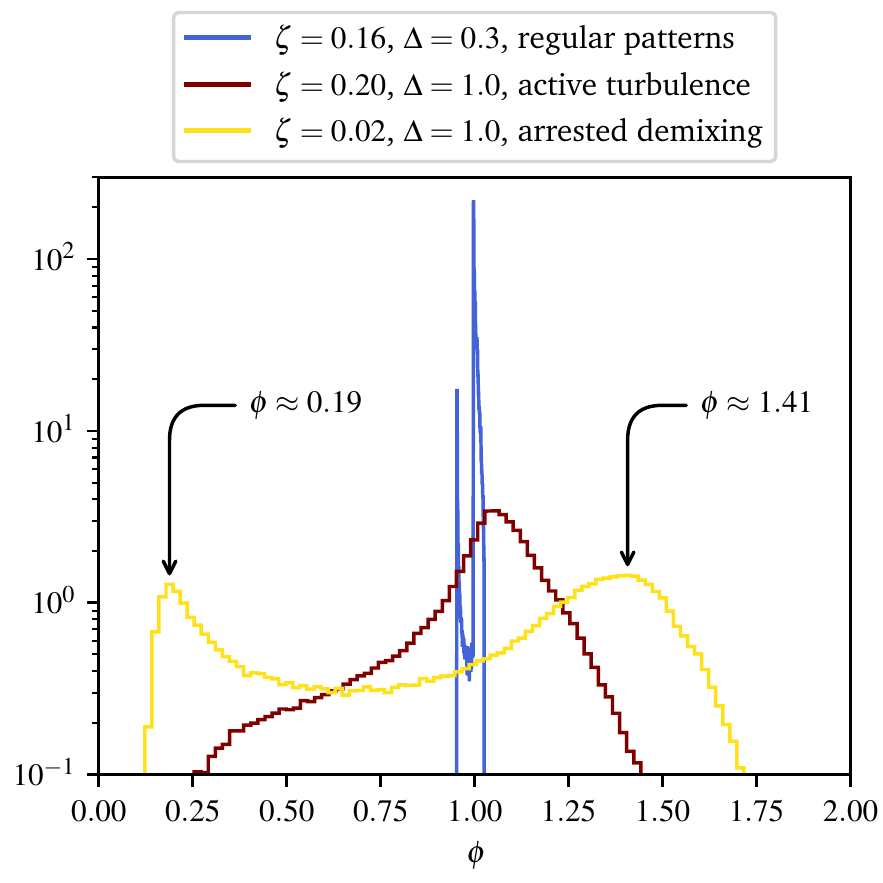}
\caption{Probability distribution functions showing frequency versus local \dom{concentration} at a grid point, for the active pattern (blue curve), the active turbulence (maroon curve) and the spontaneous phase separation (yellow curve) regimes. It can be seen that the distribution corresponding to the active turbulent regime is unimodal, while that associated with the spontaneous phase separated one is bimodal, with peaks at $\phi \approx 0.19, 1.41$, cf.\ binodal points in Fig.~\ref{fig:commontangent}.}
\label{fig5}
\end{figure}

\begin{figure}[!h]
  \includegraphics{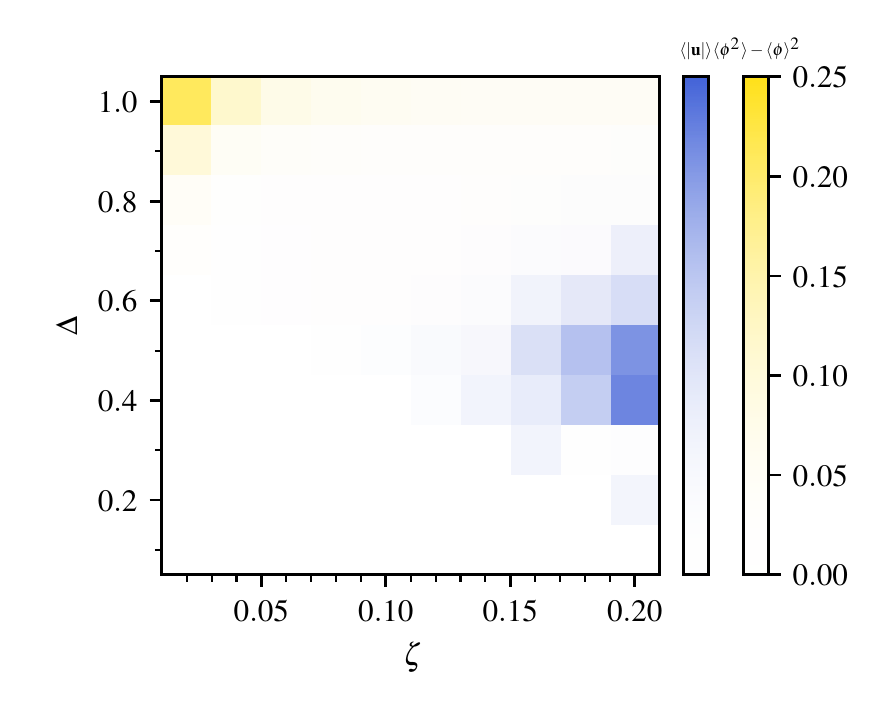}
\caption{Combined heatmap of the averaged flow magnitude, $|{\mathbf u}|$ (blue), and variance of $\phi$ (yellow), as functions of $\zeta$ and $\Delta$. The plot can be used to identify the regions of parameter space where phase separation is most prominent (upper left) or where the steady active flow is strongest (right).}
\label{fig6}
\end{figure}

\begin{figure}[!h]
  \includegraphics{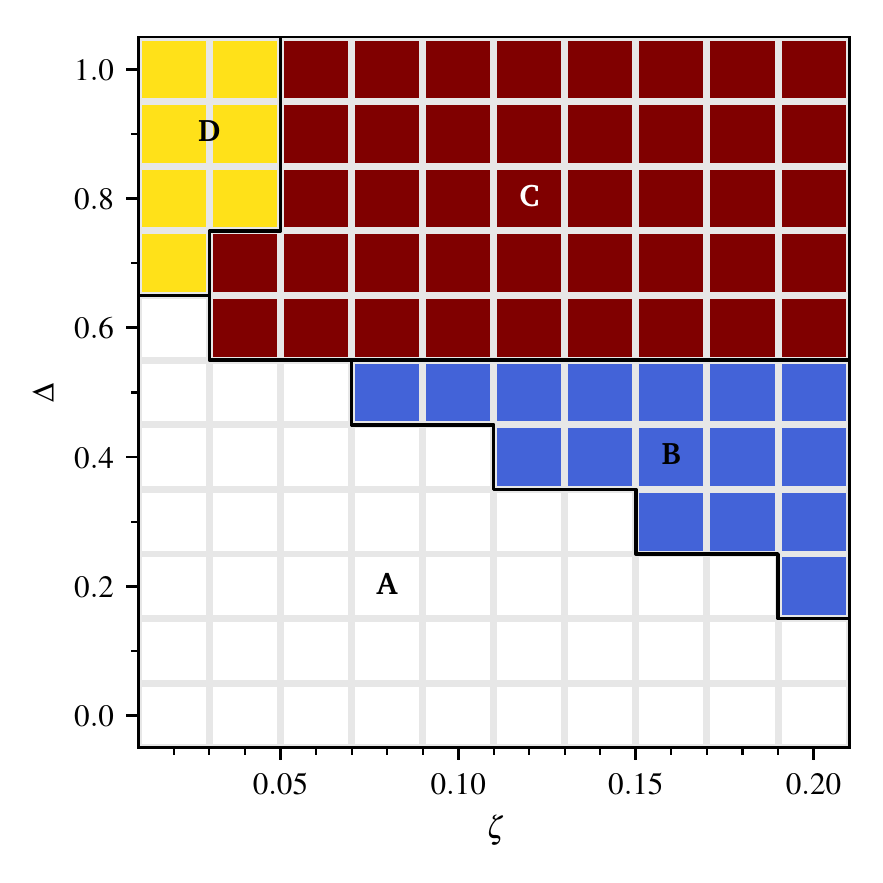}
  \caption{Phase diagram of the system in the $(\zeta,\Delta)$ plane. Each square corresponds to one simulation. Regions A--D correspond respectively to isotropic passive, active patterns, active turbulence, and spontaneous microphase separation regimes.}
\label{fig7}
\end{figure}

To quantitatively delineate the phase diagram of the system, and the boundaries between the three different dynamical regimes shown in Fig.~\ref{fig2}, we proceed as follows. We use the plot in Fig.~\ref{fig1} to identify the phases as isotropic ($\langle q\rangle \simeq 0$) or nematic ($\langle q\rangle\ne 0$). We then classify cases where the kinetic energy reaches a plateau as regular patterns. To discriminate between active turbulence and spontaneous phase separation, we look at probability distribution functions for $\phi$ (calculated over space, and averaged over time, see Fig.~\ref{fig5} for examples). If this distribution is bimodal (i.e., it has two maxima), then we classify the pattern as phase separated (Fig.~\ref{fig5}, yellow curve). A similar identification would arise from analysing the average \dom{concentration} variance and the average flow magnitude (Fig.~\ref{fig6}). Flow and \dom{concentration} variance are highest in the active pattern and spontaneous phase separation regimes respectively. The phase diagram for our inhomogeneous extensile mixture is shown in Fig.~\ref{fig7}.

The \dom{concentration} distribution plots show 
that \dom{concentration} variations are a generic feature throughout the active nematic phase. First we note that, without such \dom{concentration} variations, the spontaneous microphase separation regime could not arise.
The existence of this regime in a realisable system depends on the value of $\Delta$, which will be determined by microscopic parameters. Assuming an Onsager-like theory for \dom{composition}-dependent orientational order, we expect this could be achievable with rod-like active particles of sufficiently large aspect ratio. Note that we actually observe a bimodal distribution even in the regular pattern regime, however here the magnitude of the peaks differs by more than one order of magnitude. This corresponds to a detectable depletion of active material at the cores of long-lived defects rather than actual phase separation. A second observation is that the width of the \dom{concentration} distributions is remarkably large in the active turbulent regime, as the sampled values of $\phi$ can vary from $\sim 0$ to $\sim 2$. This range is similar to the spread observed in the spontaneous phase separation regime, although the distribution remains peaked at $\phi\simeq \phi_0=1$ for active turbulent patterns. This observation is in line with experimental microscopy of quasi-2D extensile mixtures of microtubules and molecular motors, which shows a highly inhomogeneous \dom{concentration} of active material~\cite{martinez2019}.

\if{Given the large variations in $\phi$ in the active turbulent regime, it is natural to ask whether these lead to a different qualitative behaviour with respect to active turbulence in constant-density active gels. To address this question, here we compute the velocity-velocity correlation of the active mixture. Fig.~\ref{fig8} shows that the velocity-velocity correlations differ qualitatively as $\gamma(\phi_0)$ crosses a critical value which is $\critical{\gamma}\sim 2.7$. Below $\critical{\gamma}$, the velocity correlation lengthscale --- which we can define, for instance, as the first minimum of the correlation function --- is of the order of the system size. This corresponds to a long-lived vortex which scales with the system, in line with our qualitative observations in the regular pattern regime. Above $\critical{\gamma}$, we observe a sharp reduction in correlation lengthscale, compatible with it being $\sim \sqrt{K/\zeta}$ as commonly found in active nematic turbulence. This points to the presence of two regimes in the active turbulent phase. We suggest that when the nematic phase is driven by activity (non-equilibrium nematic, where the underlying equilibrium phase is isotropic) vortices span the whole system, whereas when the underlying passive phase is nematic, then a vortices of a well-defined size appear. Interestingly, this phenomenology is qualitatively consistent with that of constant-density active gels in spite of the large density fluctuations, although a more quantitative analysis of the correlation length with $\zeta$ is required to conclude that the two regimes (inhomogeneous and constant-density active turbulence) are truly analogous.}\fi 

\subsection{Coupling-induced demixing in the passive limit}
\label{subsec:passive-demixing}

\begin{figure}[!h]
  \includegraphics[width=\columnwidth]{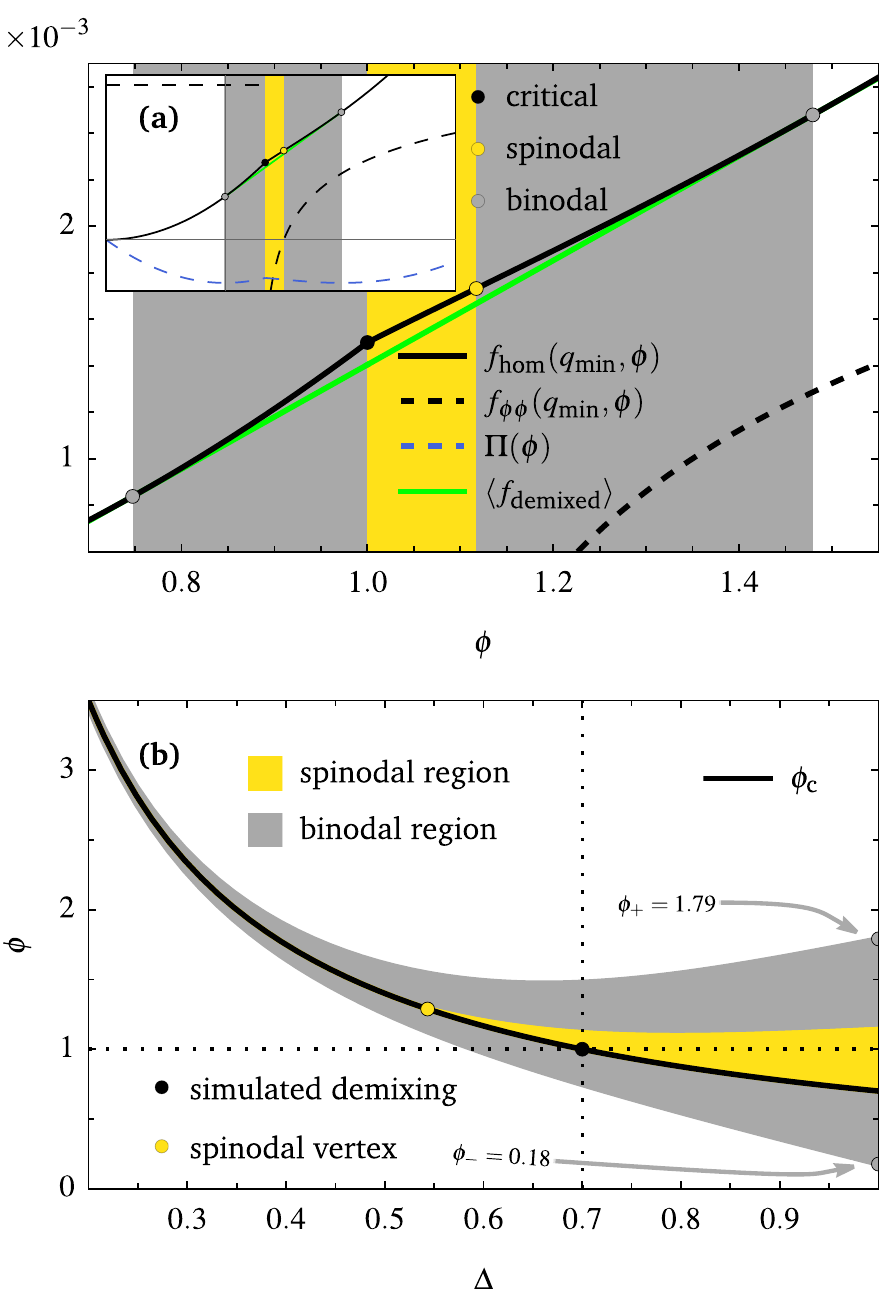}
  \caption{Coupling-induced demixing derived semi-analytically from the free energy for a uniformly aligned passive nematic phase given in Eq.~\ref{eqHomFree}. 
Panel~{\rm\bf(a)} shows the free energy, $f_\textnormal{hom} \left( \minimizer{q}(\phi), \phi \right)$, which, for a given value of $\phi$, is minimized by $\minimizer{q}(\phi)$ (see Section~\ref{subsec:passive-demixing}). The second derivative with respect to $\phi$ is shown with a dashed curve, and is negative in the spinodal region, highlighted in yellow. Here the system is linearly unstable to demixing into regions of high \dom{($\phi_+$)} and low \dom{($\phi_-$) concentration}, the values of which are given by the respective binodal points, shown in grey. These points are found by the \dom{common} tangent construction illustrated by the green line, and share the same chemical potential and pressure ($\Pi$, dashed in blue). 
In this plot $\Delta$ is fixed at 0.7, and the presence of the lower spinodal point at $\phi = 1$ \dom{compares well} with our simulation results, as illustrated in Fig.~\ref{fig7}. 
Panel~{\rm\bf(b)} shows the $\phi$--$\Delta$ plane, which is a phase diagram derived from the loci of the spinodal and binodal points illustrated in panel~{\rm\bf(a)}, considered as we vary $\phi$ and $\Delta$. The shaded spinodal and binodal regions are bounded by these loci, and show the parameter combinations for which the system is unstable to phase decomposition. 
\dom{Note that all $\phi$--axis values signify the spatial average over the system, which is dropped from the notation for readability.}
}
\label{fig:commontangent}
\end{figure}

To shed more light on the physics underlying the occurrence of our spontaneous microphase separated regime, it is instructive to consider the passive limit of our nematic mixture, where the behaviour is solely determined by thermodynamic considerations. In this case, the free energy is minimised in equilibrium, and a full phase diagram can be computed semi-analytically using the common tangent construction (Fig.~\ref{fig:commontangent}). 

Recall that our model employs both conserved (composition $\phi$) and non-conserved (nematic tensor $Q_{\alpha\beta}$) order parameters, which are mutually coupled through the free energy given in Eq.~\ref{eqBulkFree}. The resulting mixture is an example of a lyotropic liquid crystal, such as that considered in~\cite{matsuyama2002}, albeit with an important distinction. Specifically, in the limit of no coupling ($\Delta=0$ in Eq.~\ref{eqCoupling}), the system we consider exhibits no phase separation and mixes freely. In what follows, therefore, we shall see that passive phase separation is driven only by the coupling.

To find the equilibrium state, we consider a uniformly aligned (homogeneous) nematic phase, so that the elastic term contribution in Eq.~\ref{eqBulkFree} vanishes and the total free energy can be written in terms of the magnitude of nematic order, $q$, and of $\phi$ as follows,
\begin{align}
  f_\textnormal{hom} \dom{\left( q, \phi \right)}
  &=
  \frac{A_0}{3} \left(1 - \frac{\gamma(\phi)} {3} \right) q^2
  - \frac{2 A_0 \gamma(\phi)}{27} q^3
  + \frac{A_0 \gamma(\phi)}{9} q^4 + \frac{a}{2}\phi^2
  .
\label{eqHomFree}
\end{align}

\dom{This homogeneous state has everywhere the same value of the conserved composition. Thus, for each point $(x, y)$ in the system, $\phi(x, y) = \frac{\int\ddsp{x} \ddsp{y} \phi(x, y)}{\int\ddsp{x} \ddsp{y}} \equiv \mean{\phi}$. The magnitude of order that minimizes the total free energy for this given homogeneous value of $\phi$ is}
\begin{align}
  \minimizer{q}(\phi)
  &=
  \begin{cases}
    0,
      &\text{ if } \gamma \leq \critical{\gamma}\\
    \tfrac14 \left( 1 + \sqrt{9 - 24/\gamma(\phi)} \right),
      &\text{ if } \gamma \geq \critical{\gamma} \dom{.}
  \end{cases}
\end{align}

\dom{If the homogeneous state is stable against phase decomposition into regions of high and low concentration, then there is no configuration for which $\mean{f}$, the spatial average of the free energy density, is less than the value of $f_\textnormal{hom} \left( \minimizer{q}(\phi), \phi \right)$. However, as the common tangent construction illustrated in Fig.~\ref{fig:commontangent}a shows, for each homogeneous configuration lying between the binodal points, there exists a corresponding inhomogeneous configuration with the same average composition, $\mean{\phi}$, but a lower average free energy density, $\mean{f_\textnormal{demixed}}$, the value of which is given by the tangent line itself. Moreover, for sufficiently large $\Delta$ and certain values of $\phi$, we find that $f_\textnormal{hom} \left( \minimizer{q}(\phi), \phi \right)$ is a non-convex function. In this spinodal region, we therefore expect that the nematic mixture is linearly unstable to phase separation. By comparison, in the limit of vanishing activity, our simulation found phase decomposition in the parameter range $\mean{\phi}=1$, $\Delta>0.7$ (see, for instance, the left border of Fig.~\ref{fig7}). This simulated demixing is highlighted in Fig.~\ref{fig:commontangent}b, and signifies good agreement with our semi-analytic thermodynamic prediction}.

In greater detail, the spinodal region --- where the system is linearly unstable to phase separation, for any perturbation however small --- is precisely where $f_\textnormal{hom} \left( \minimizer{q}(\phi), \phi \right)$ is concave down. Thus it is bounded by the inflexion points, where the second derivative of $f_\textnormal{hom}$ with respect to $\phi$, $f_{\phi\phi} \left( \minimizer{q}(\phi), \phi \right)$, crosses the $\phi$ axis in Fig.~\ref{fig:commontangent}a. The lower spinodal point is simply $\critical{\phi}=(\critical{\gamma} - \gamma_0) / \Delta$, whereas the upper spinodal point, $\spinodal{\phi}$, was found numerically as the only real root of a seventh order polynomial.

This effective free energy also gives, by a \dom{common} tangent construction, the binodal points, namely the values of the coexisting \dom{concentrations} of the phase separated systems. Specifically, as coexisting phases must share a common chemical potential ($\mu = f_\phi \left( \minimizer{q}(\phi), \phi \right)$) and a common pressure ($\Pi = f_\textnormal{hom} \left( \minimizer{q}(\phi), \phi \right) - \mu \phi$), we have a system of two equations in two unknowns. The two unknowns in question are the two \dom{concentrations, $\phi_-$ and $\phi_+$,} for which a straight line tangentially touches the curve $f_\textnormal{hom} \left( \minimizer{q}(\phi), \phi \right)$ exactly twice, and were found numerically. \dom{For values of $\mean{\phi}$ between the binodal points, this tangent represents the average free energy density of the phase separated system, which is visibly lower than the free energy density of the homogeneous system. Thus $\mean{f_\textnormal{demixed}} < f_\textnormal{hom} \left( \minimizer{q}(\phi), \phi \right)$ for $\phi_- < \mean{\phi} <\phi_+$.}

The loci of these spinodal and binodal points as we vary $\phi$ and $\Delta$ are the \dom{curv}es plotted in the $\phi$--$\Delta$ plane in Fig.~\ref{fig:commontangent}b. Here we can see that, in line with our simulation \dom{results shown in Fig.~\ref{fig7}}, there is a spinodal point at $\Delta=0.7$ for $\mean{\phi}=1$.

We stress again that the analysis in this section refers to the $\zeta=0$ passive limit of our mixture. In the active case, macroscopic phase separation is generically arrested by the activity-induced spontaneous flow. This leads to microphase separation at steady state, \dom{similarly to what has been observed in sheared passive binary mixtures~\cite{stansell2006,2004.Viscoelastic-phase-separation-shear-flow.Imaeda-Furukawa-Onuki,taniguchi1996}} or in active model H~\cite{tiribocchi2015}.

Quantitatively, our theory predicts a spinodal point at $\Delta=0.7$ for $\mean{\phi}=1$, which \dom{compares well} with our simulation results (Fig.~\ref{fig7}). The predicted binodal points for $\Delta = 1$, lie at $\phi=0.18$ and $\phi=1.79$ --- \dom{these values should be compared} with the histogram peaks shown in Fig.~\ref{fig5}, viz.\ $\phi=0.19$ and $\phi=1.41$ --- the lower value of the latter peak is likely due to the active flow which drives the system away from thermodynamic equilibrium (assumed in Fig.~\ref{fig:commontangent}). \davide{Note our simulations only find phase separation in the spinodal region in Fig.~\ref{fig:commontangent}b, as the initial perturbation of the homogeneous state are smaller than the scale required for nucleation.}


\section{Discussion and Conclusions}

In summary, we have used computer simulations to study the hydrodynamics of an inhomogeneous active nematic gel. With respect to conventional models for active gels, which only consider the velocity field and ${\bf Q}$ tensor, our theory also allows for the time evolution of the active matter \dom{concentration} $\phi$. Previous work has shown by a linear stability analysis that \dom{compositional} fluctuations are irrelevant for the physics of the ``generic instability'' of active gels~\cite{baskaran2009}, which stands for the transition between the passive (quiescent) and the active (spontaneously flowing) phase. It has however remained unclear what their role is deep in the active phase, where nonlinearities are important; shedding light on this issue has been the focus of our current work. 

Our main result is the quantitative characterisation of the phase diagram of inhomogeneous active nematic (Fig.~\ref{fig7}). We have found that there are three regimes with distinct emergent behaviour in the active phase. First, close to the transition between the passive isotropic and active nematic phase, there are regular patterns typically composed of self-assembled rotating spirals. Second, deeper in the active nematic phase there is an active turbulent regime featuring chaotic dynamics of vortices and half-integer nematic defects. Third, for low activity and large enough nematic tendency ($\Delta$ in our phase diagram in Fig.~\ref{fig7}), we find spontaneous phase separation into active and passive domains. This latter phase separation is arrested by the active flow, so that domains do not coarsen past a typical size, which decreases with increasing activity.

The regular spiral/vortex patterns we find are reminiscent of those observed with polar active gels in the ordered phase~\cite{kruse2004,elgeti2011}. While polar nematics can only exhibit defects with integer topological charge, defects in active (apolar) nematics normally have half-integer topological charge, so it is non-trivial that close to the passive-active transitions we observe spirals, whose topological charge is $+1$.~\footnote{Note that in our geometry integration of the topological charge density, defined as in~\cite{blow2014}, is conserved and equal to $0$.} Notwithstanding this qualitative resemblance, the patterns we observe also have a non-trivial spatiotemporal dynamics (Suppl.\ Movie~\dom{2}). The core of our spirals are also associated with notable \dom{concentration} minima, or voids, which arise because of the coupling between nematic order and \dom{composition} in the free energy of the system. The mechanism responsible for this coupling is the same that drives inert colloidal particles or isotropic droplets (with no anchoring on their surface) to the defect cores or disclinations in passive liquid crystals~\cite{ravnik2011,foffano2019}.

The chaotic, active turbulent regime we find for sufficiently large $\zeta$ is an analog of the regime \dom{of the same name} in active gels \dom{of uniform composition}~\cite{thampi2013,linkmann2019,carenza2020,alert2020}. An important feature of this regime in our simulations, though, is that there are very large \dom{compositional} fluctuations (Fig.~\ref{fig5}a). These are qualitatively in line with experimental observations of active turbulence in microtubule--motor mixtures, which show substantial inhomogeneities in microtubule \dom{concentration} over a sample~\cite{sanchez2012,martinez2019}. Whether such \dom{concentration} variations lead to a fundamental change in the scaling properties of active turbulence is an open question which we believe deserves further investigation, for instance by a quantitative detailed analysis of the scaling of velocity--velocity correlations~\cite{stenhammar2017,bardfalvy2019,skultety2020}.

Regarding the spontaneous microphase separated regime, this is notable especially because there is no term in the free energy density $f_{\phi}$ which explicitly favours demixing. In other words, in the absence of $f_{\rm LC}$ the system would remain uniform. Phase separation arises due to the coupling between \dom{composition} and order, measured by the parameter $\Delta$. In this sense, phase separation is not driven by activity but rather thermodynamically, and indeed it can be explained with a theoretical discussion of the free energy in the passive limit (Fig.~\ref{fig:commontangent}). In simulations we observe a microphase separated pattern rather than macroscopic phase separation, as the active flow arrests coarsening, and controls the size of the steady-state domains observed at late times, 
similarly to the case of active model H~\cite{tiribocchi2015}. While experimental realisation of active nematics have shown plenty of instances of active turbulence~\cite{sanchez2012,martinez2019}, the spontaneous microphase separation regime appears to not have been found in the lab yet. Our model suggests that the most promising avenue to realise this regime experimentally is to control the \dom{composition}-order coupling $\Delta$. The latter may be estimated by monitoring how the isotropic--nematic transition point depends on the \dom{concentration} of nematogenic particles (for instance, microtubules) in the passive limit of no activity. 

\if{\begin{figure}[!h]
  \includegraphics[width=0.49\columnwidth]{snapshot_defects_director.png}
  \includegraphics[width=0.49\columnwidth]{snapshot_defects_flow.png}
  \includegraphics[width=0.49\columnwidth]{snapshot_rafts_director.png}
  \includegraphics[width=0.49\columnwidth]{snapshot_rafts_flow.png}
  \caption{Here we should show the snapshots (Q, $\phi$, flow) for the raft regime with compressibility.}
\label{fig9}
\end{figure}
}\fi

Looking ahead, we can suggest a few directions in which our work can be carried forward. First, it would be of interest to understand from a more fundamental point of view the universal properties of the dynamical regimes we have identified. For instance, one could quantify the dependence of vortex correlation length and pattern size on physical parameters, and the scaling of the power spectra of the kinetic energy. This would allow to clarify the important theoretical question of whether or not inhomogeneous active turbulence is in the same universality class as turbulence in active gels \dom{of uniform composition}. Second, from the experimental point of view, it would be desirable to compare more quantitatively \dom{concentration} distributions in active turbulence with those predicted by our simulations. Third, with regards to computer simulations, it would be highly interesting to explore the phase behaviour and dynamics of inhomogeneous active nematics in 3D, comparing and contrasting it with the one found here in 2D.

\section{Acknowledgement}

We thank S. Samatas for useful discussions. 
For the purpose of open access, the authors have applied a Creative Commons Attribution (CC BY) licence to any Author Accepted Manuscript version arising from this submission. This work was partially funded by EPSRC (grant number EP/V048198/1). We would also like to thank EPSRC for the computational time made available on the ARCHER2 UK National Supercomputing Service (https://www.archer2.ac.uk).

\bibliography{densitydependentactivegel} 
\bibliographystyle{rsc} 
\end{document}


\title{Active turbulence and spontaneous phase separation in inhomogeneous extensile active gels\\Supplemental Material}

\author{Renato Assante}
\affiliation{SUPA, School of Physics and Astronomy, The University of Edinburgh, James Clerk Maxwell Building, Peter Guthrie Tait Road, Edinburgh, EH9 3FD, United Kingdom}

\author{Dom Corbett}
\affiliation{SUPA, School of Physics and Astronomy, The University of Edinburgh, James Clerk Maxwell Building, Peter Guthrie Tait Road, Edinburgh, EH9 3FD, United Kingdom}

\author{Davide Marenduzzo}
\affiliation{SUPA, School of Physics and Astronomy, The University of Edinburgh, James Clerk Maxwell Building, Peter Guthrie Tait Road, Edinburgh, EH9 3FD, United Kingdom}

\author{Alexander Morozov}
\affiliation{SUPA, School of Physics and Astronomy, The University of Edinburgh, James Clerk Maxwell Building, Peter Guthrie Tait Road, Edinburgh, EH9 3FD, United Kingdom}

\date{\today}

\maketitle

\section{Characteristic domain scale in microphase separation}
\label{sec:lengthscalevstime}

To support the claim of microphase separation for the low activity and large nematic tendency, the coarsening of domains was quantified by examining the time-evolution of the characteristic domain scale, $L(t)$. This was computed as $2 \pi$ times the inverse of the first moment of the static structure factor $S(k, t) = \mean{\delta\phi(k, t) \delta\phi(k, t)}$, where $\delta\phi = \phi - \phi_0$ and angular brackets denote here the ensemble average, so that
\begin{align}
  L(t)
  &=
  2 \pi
  \left [
  \frac{
  \int_\mathcal{K} \ddsp{k} S(k, t) k
  }{
  \int_\mathcal{K} \ddsp{k} S(k, t)
  }
  \right]^{-1}
  ,
\end{align}
where $\mathcal{K}$ is the set of modes implemented within the respective run of our in-house spectral code. 

\begin{figure}[!h]
  \includegraphics{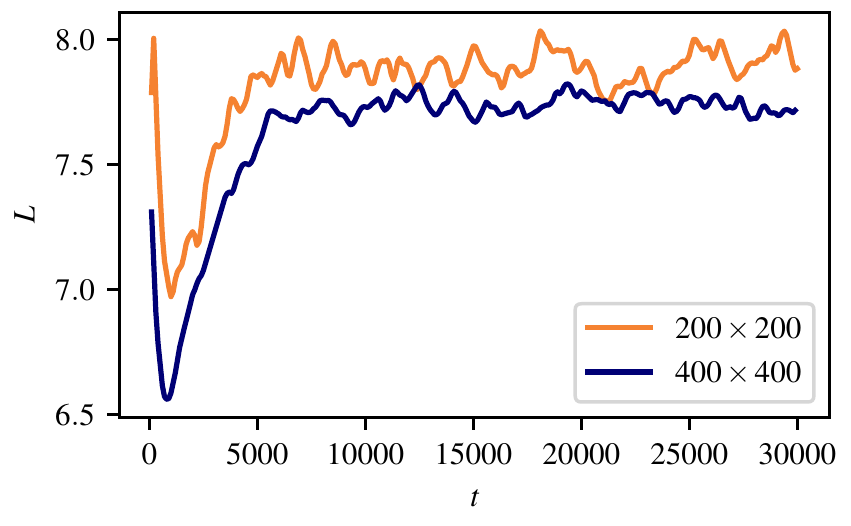}
  \caption{Time-evolution of characteristic domain scale for microphase separated regime as described in section~\ref{sec:lengthscalevstime}, plotted for simulated system sizes (side of simulation box, $H$) of $200$ and $400$ respectively.}
  \label{fig:lengthscalevstime}
\end{figure}

\begin{figure}[!h]
  \includegraphics{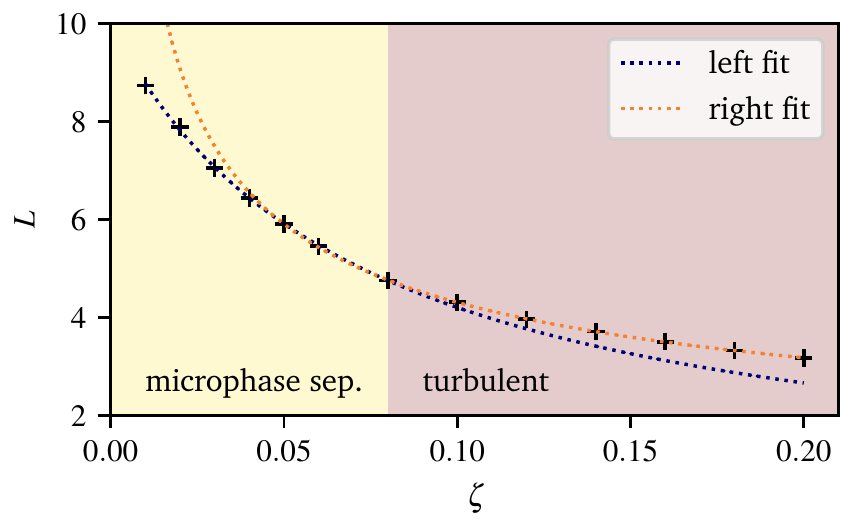}
  \caption{Characteristic domain scale for a range of activities crossing the boundary between the microphase separated (left) and turbulent (right) regimes, for $\Delta=1$. The different behaviour is highlighted by shading corresponding to each regime and a fit to the respective set of points, viz.\ $(1.4 \zeta + 0.1)^{-1}$ and $1.2 \zeta^{-1/2} + 0.45$.}
  \label{fig:lengthscale-vs-activity}
\end{figure}

As Fig.~\ref{fig:lengthscalevstime} shows, for a given activity the characteristic domain scale increases until reaching a plateau indicative of arrested coarsening. The plateaus for different system sizes are very similar, indicating that the late-time characteristic domain scale does not scale with system size.
Fig.~\ref{fig:lengthscale-vs-activity} additionally shows that the characteristic domain scale decreases with activity. Our data suggest that the functional form of the decay may differ in the microphase separated and active turbulent regime. It would be of interest to explore this issue further, for instance via finite size scaling.

\section{Incompressibility and Navier--Stokes solvers}

In this Section we discuss a technical point, which is of relevance to the high-activity behaviour of our system. While the {concentration} of active matter varies, the overall fluid is incompressible: in other words, the total density of active matter and underlying solvent is constant. Incompressibility is important especially for high activity and intermediate $\Delta$ --- i.e., in the bottom right region in the phase diagram in Fig.~5, which is challenging to characterise accurately. There we typically find active turbulent patterns with vortex correlation length similar to the system size. We found that these are replaced by an apparent phase separation into high and low-{concentration} patches if the Navier--Stokes solver allows for fluid compressibility and the Mach number is not sufficiently small, which occurs when using, for instance, hybrid Lattice Boltzmann simulations at large activity (in our case, we found this for $\zeta>0.1$). 
These states are artifacts in the current model, and using a purely incompressible and 2D fluid they are not found, in line with simulations with tracers in turbulent flows which only show aggregation in compressible fluids, or due to inertial effects~\cite{falkovich2001,das2000}. It would however be of interest to see whether these patterns may be recovered in thin active matter films, which can behave effectively as compressible fluids~\cite{voituriez2006}. 
\begin{figure}[p]
  \includegraphics[width=\columnwidth]{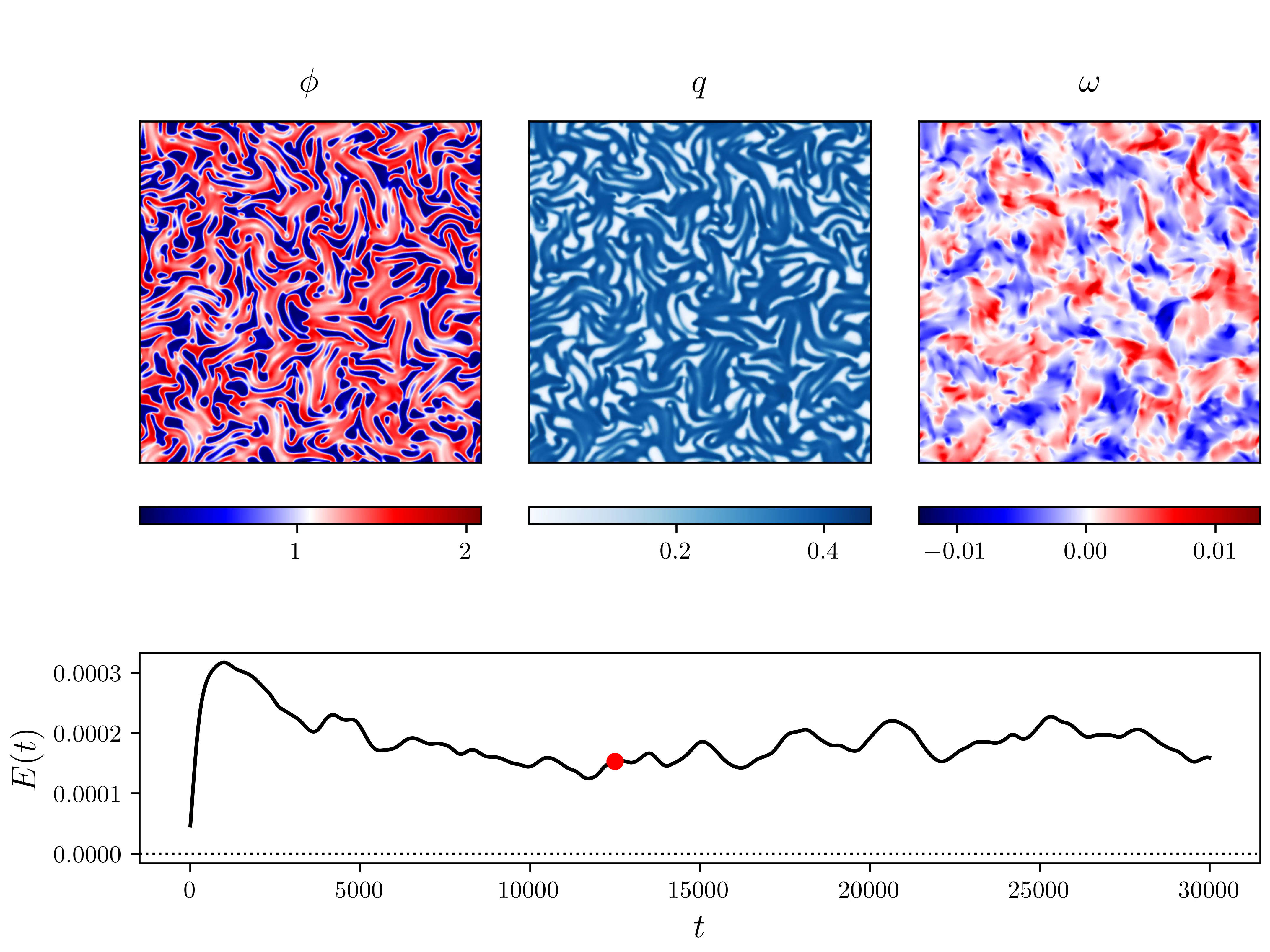}
  \caption{Initial snapshot of Suppl. Movie 1.}
\end{figure}

\begin{figure}[p]
  \includegraphics[width=\columnwidth]{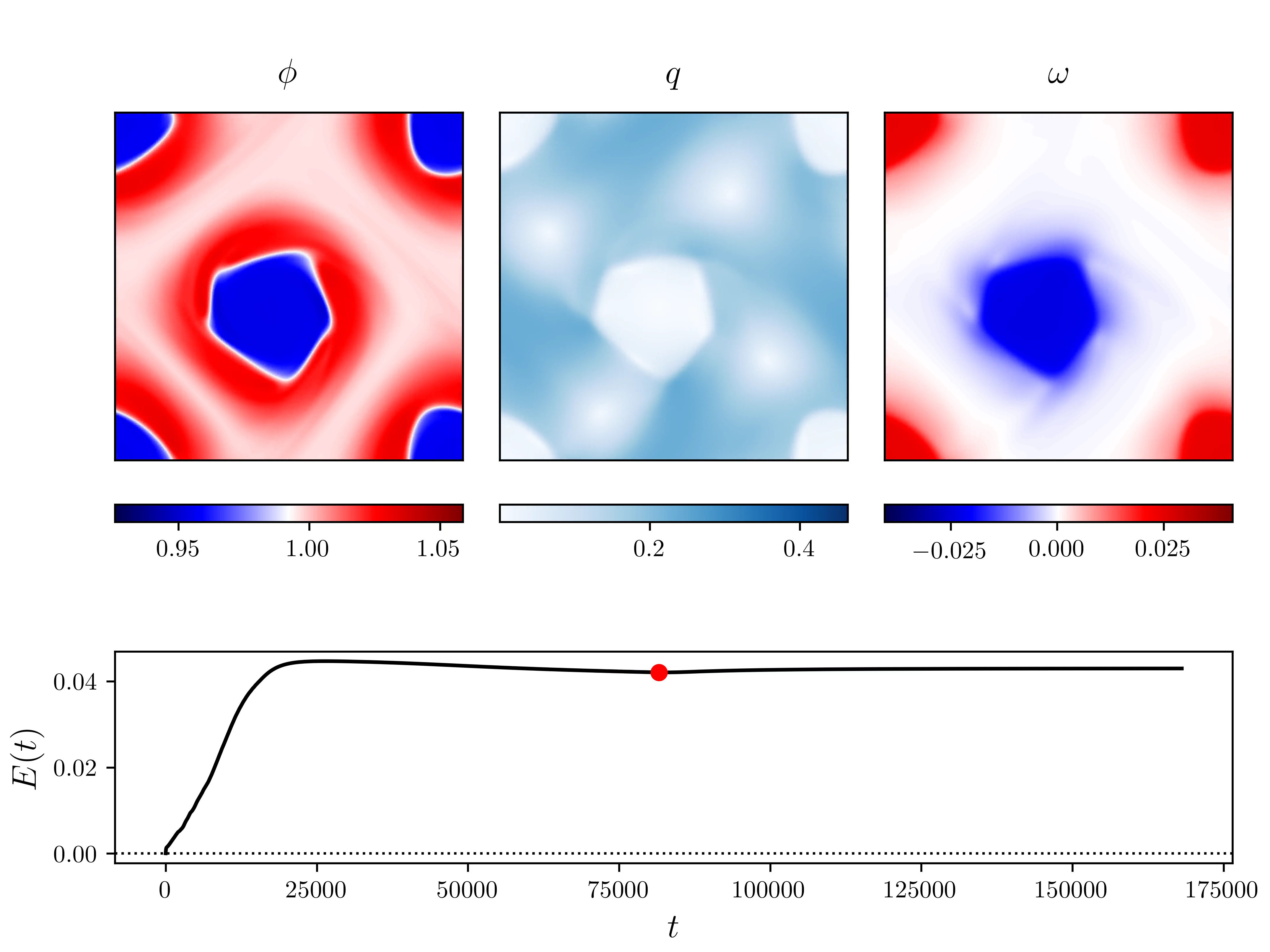}
  \caption{Initial snapshot of Suppl. Movie 2.}
\end{figure}

\begin{figure}[p]
  \includegraphics[width=\columnwidth]{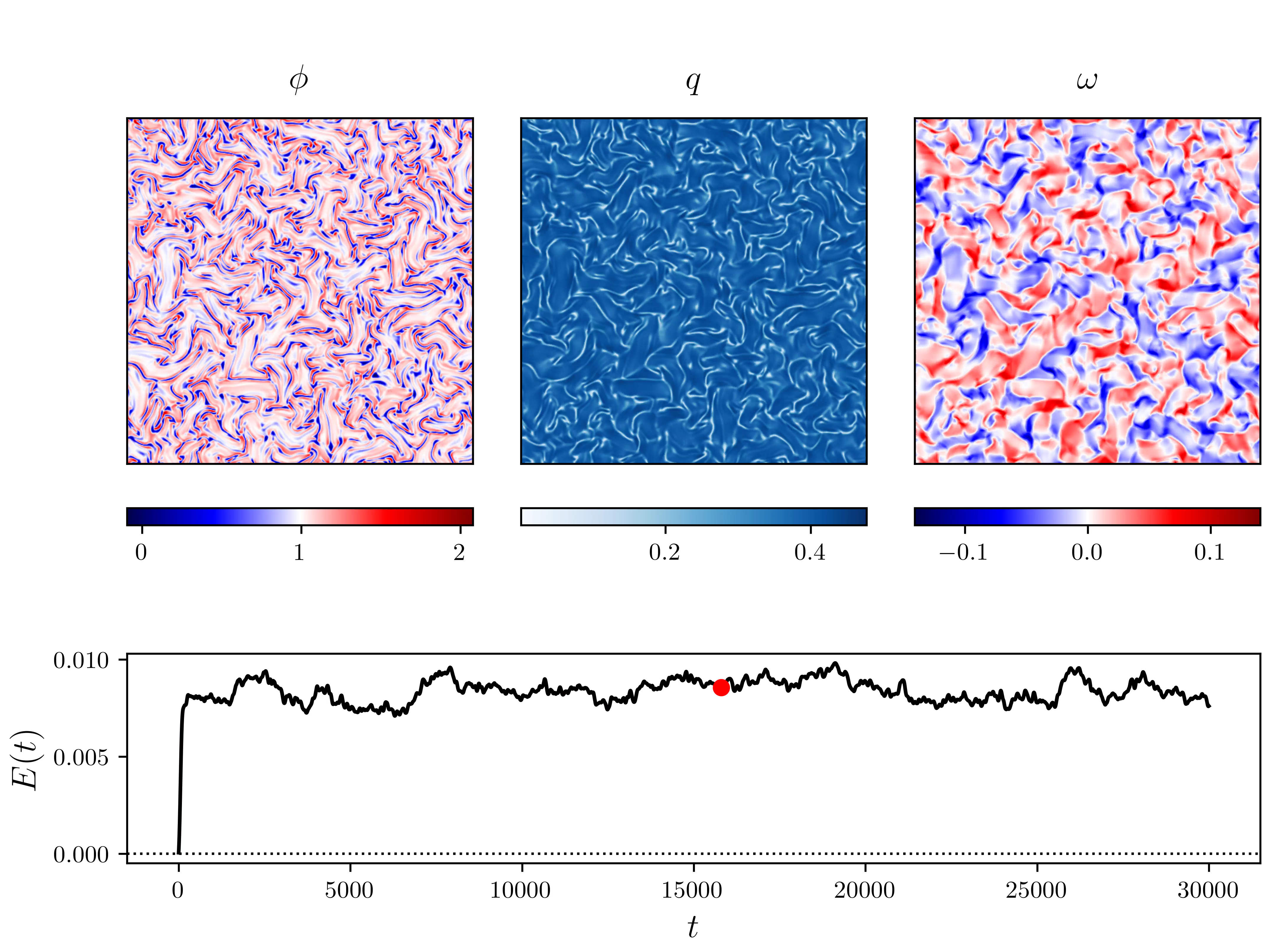}
  \caption{Initial snapshot of Suppl. Movie 3.}
\end{figure}

\section{Supplemental Movies}

Captions for the Supplemental Movies are given below. Corresponding snapshots are shown in Figs.~S1-S3.

{\bf Suppl. Movie 1:} this movie shows the dynamics corresponding to the microphase separated regime. The graph at the bottom shows the time evolution of the kinetic energy.

{\bf Suppl. Movie 2:} same as Suppl. Movie 1, but for the regimes where patterns are observed.

{\bf Suppl. Movie 3:} same as Suppl. Movie 1, but for the active turbulence regimes.


\bibliography{densitydependentactivegel}